\documentclass[final]{IEEEtran}
\pdfoutput=1
\usepackage{amsthm,amssymb,graphicx,multirow,amsmath,color,amsfonts}
\usepackage[update,prepend]{epstopdf}
\usepackage[noadjust]{cite}
\usepackage[latin1]{inputenc}
\usepackage{tikz}
\usepackage{bbm} 
\usepackage{pdfpages}
\usepackage{tabulary}
\usepackage{multirow}
\usepackage{comment}

\def\nb0{{\mathbf{0}}}
\def\nb1{{\mathbf{1}}}

\newtheorem{lemma}{Lemma}

\newtheorem{definition}{Definition}

\newtheorem{theorem}{Theorem}

\newtheorem{remark}{Remark}

\allowdisplaybreaks 

\begin{document}
\graphicspath{{./Figures/}}
\title{
Stochastic Geometry-based Comparison of Secrecy Enhancement Techniques in D2D Networks}
\author{
Mustafa A. Kishk and Harpreet S. Dhillon
\thanks{The authors are with Wireless@VT, Department of ECE, Virginia Tech, Blacksburg, VA. Email: \{mkishk, hdhillon\}@vt.edu. The support of the U.S. NSF (Grants CCF-1464293 and CNS-1617896) is gratefully acknowledged.} 
\vspace{-9mm}}

\maketitle

\begin{abstract}
This letter presents a performance comparison of two popular secrecy enhancement techniques in wireless networks: (i) {\em creating guard zones} by restricting transmissions of legitimate transmitters whenever any eavesdropper is detected in their vicinity, and (ii) {\em adding artificial noise} to the confidential messages to make it difficult for the eavesdroppers to decode them. Focusing on a noise-limited regime, we use tools from stochastic geometry to derive the secrecy outage probability at the eavesdroppers as well as the coverage probability at the legitimate users for both these techniques. Using these results, we derive a threshold on the density of the eavesdroppers below which no secrecy enhancing technique is required to ensure a target secrecy outage probability. For eavesdropper densities above this threshold, we concretely characterize the regimes in which each technique outperforms the other. Our results demonstrate that guard zone technique is better when the distances between the transmitters and their legitimate receivers are higher than a certain threshold.%
\end{abstract}
\vspace{-4mm}
\section{Introduction} \label{sec:intro}
Owing to the broadcast nature of wireless networks, physical layer security techniques are necessary to preserve confidentiality of the transmitted messages~\cite{5934342,zhou2010secure,7247765,6259796}. Two popular secrecy enhancing techniques that have been investigated in the literature are: (i) {\em creating guard zones} by restricting transmissions of the legitimate transmitters whenever eavesdroppers are detected in their vicinity~\cite{5934342}, and (ii) {\em adding artificial noise} to the confidential messages to make it difficult for the eavesdroppers to decode them~\cite{zhou2010secure}. Despite the attention received by these techniques, to the best of our knowledge their explicit system-level performance comparison is still an open problem, which is the main focus of this letter. 

The system-level analysis of wireless networks usually requires averaging the performance metric of interest over all possible topologies of the network. While this has traditionally been performed through Monte-Carlo trials, stochastic geometry has recently emerged as an attractive analytic alternative due to its remarkable tractability~\cite{AndGupJ2016}. In fact, stochastic geometry has also gained popularity in the past few years for the system-level analysis of D2D networks, e.g., see~\cite{7073589,7056528,7446343}, as well as physical layer security, e.g., see~\cite{5934342,geraci2014physical}. In particular,~\cite{5934342} quantified the loss in system throughput that results from ensuring a specific level of secrecy in decentralized wireless networks. Similarly,~\cite{geraci2014physical} studied physical layer security in downlink cellular networks assuming the downlink messages meant for each user can be eavesdropped by all other users (both intra- and inter-cell) in the network. 

In this letter, we will use tools from stochastic geometry for the comparison of secrecy enhancing techniques. In addition to the two techniques introduced already, namely, creating guard zones and adding artificial noise, there are two other techniques usually considered in the literature: (i) {\em protected zones}, and (ii) {\em beamforming}. Protected zones are similar to the guard zones defined earlier in this section, except that they are guaranteed to be free of eavesdroppers (physically enforced)~\cite{7247765}. If one assumes multi-antenna nodes, beamforming is also an attractive solution for enhancing secrecy~\cite{6259796}. In this letter, we consider a system with single-antenna nodes in which we do not have control over the physical removal of eavesdroppers, as a result of which we focus only on the first two techniques (guard zones and artificial noise).

\textit{Contributions}. We consider a device-to-device (D2D) network that coexists with a network of eavesdroppers modeled by an independent Poisson point process (PPP). For this setup, focusing on the noise-limited regime, we first derive the secrecy outage probability at the eavesdroppers and coverage probability at the legitimate receivers for the two secrecy enhancement techniques considered in this letter. Using these results, we characterize the maximum density of eavesdroppers below which no secrecy enhancing technique is required to ensure the target secrecy outage probability. For eavesdropper densities above this threshold, we concretely characterize the regimes in which a given technique outperforms the other, which leads to useful system design insights.%
\vspace{-3mm}
\section{System Model} \label{sec:SysMod}
Focusing on the noise-limited regime, we consider a primary D2D link coexisting with a secondary network of potential eavesdroppers modeled as an independent PPP $\Phi_{e}\equiv\{y_{i}\}\subset\mathbb{R}^2$ with density $\lambda_{e}$. The D2D link is formed by a primary transmitter (PT) located at the origin and a primary receiver (PR) located at a fixed distance $d$ from the PT (at an arbitrary angle from the origin). We assume independent Rayleigh fading on all wireless links. The PT is assumed to transmit at a fixed power $P_{t}$. For this setup, the received power at the PR associated with the PT is $P_{t}h\|d\|^{-\alpha}$, where $h\sim \exp(1)$ models Rayleigh fading, $\|d\|^{-\alpha}$ is the standard power-law path-loss with exponent $\alpha > 2$. Similarly, the received power at an arbitrary eavesdropper located at $y \in \Phi_{e}$ from the PT is $P_{\rm t}g_{y}\|y\|^{-\alpha}$, where $g_{y}\sim \exp(1)$ models Rayleigh fading. For the secrecy outage analysis, we will need to analyze the performance of eavesdroppers that have the best chance of decoding the messages from the PT.  The location of this {\em strongest} eavesdropper corresponding to the PT is:
\begin{align}
\label{1}
{\small y^*=\arg\underset{y\in\Phi_{e}}{\max}\ g_{y}\|y\|^{-\alpha}}.
\end{align}   
According to Wyner encoding scheme~\cite{6772207}, the transmitter chooses a rate of codeword transmission $\mathcal{C}_t$ and a rate of confidential message transmission $\mathcal{C}_s$. The rate difference, $\mathcal{C}_e=\mathcal{C}_t-\mathcal{C}_s$, represents the cost of securing the confidential message where perfect secrecy is achieved as long as mutual information between the PT and the eavesdropper is lower than $\mathcal{C}_e$. Please refer to Sec. II-B in~\cite{4802331} for further details on Wyner encoding scheme. As is usually the case in the literature, e.g., see~\cite{7247765}, we assume a noise-limited scenario for analytical tractability. Therefore, in order to ensure successful decoding at the PR, we need to satisfy the condition $\log_2(1+{\rm SNR_P})\geq\mathcal{C}_t$, where ${\rm SNR_{P}}$ is the SNR achieved at the PR. On the other hand, to ensure perfect secrecy we require $\log_2(1+{\rm SNR_S})\leq\mathcal{C}_e$ at the eavesdropper located at $y^{*}$, where ${\rm SNR_{S}}$ is the SNR at the eavesdropper. Equivalently, we can define two thresholds $\beta_t=2^{\mathcal{C}_t}-1$ and $\beta_e=2^{\mathcal{C}_e}-1$ on ${\rm SNR_P}$ and ${\rm SNR_S}$ respectively. For this setup, we now define two main performance metrics that will be used in this work.%
\begin{definition}[Coverage probability] \label{def:1}
The SNR coverage probability at the PR is defined as 
\begin{align}
P_{\rm cov}=\mathbb{P}({\rm SNR_P}\geq\beta_t,\delta_a=1),
\end{align}
where $\delta_a=1$ if the PT is transmitting information (referred to as an active PT), and $\delta_a=0$ otherwise. 
\end{definition}
\begin{definition}[Secure communication probability~\cite{5701754}] \label{def:2}
It is the probability of perfect secrecy of the confidential message from the PT (conditioned on the fact that PT is active): 
\begin{align}
P_{\rm sec}=\mathbb{P}({\rm SNR_S}\leq\beta_e|\delta_a=1)
\end{align}
\end{definition}
Our main objective is to maximize the SNR coverage probability at the PR while ensuring that the secure communication probability is above a predefined threshold $\epsilon$. 
\vspace{-4mm}
\section{Secrecy Enhancing Techniques}
\subsection{Guard Zone Technique}
In this technique, a given PT is allowed to transmit confidential messages to its paired PR only if there are no eavesdroppers in a circular {\em guard zone} of radius $r_g$ around it. Therefore, the probability that the PT is active is: 
\begin{align}
\label{2}
P_{\rm active}=\mathbb{P}(\delta_a=1)=\mathbb{P}(\mathcal{N}(\mathcal{B}(o,r_g))=0)=e^{-\lambda_e\pi r_g^2},
\end{align}
where $\mathcal{N}(\mathcal{B}(o,r_g))$ is the number of eavesdroppers inside a ball of radius $r_g$ centered at the origin. Owing to the independence of ${\rm SNR_P}$ and $\mathcal{N}(\mathcal{B}(o,r_g))$, the SNR coverage probability for the D2D link is defined as: 
\begin{align}
\label{3}
P_{\rm cov}^{GZ}&=P_{\rm active}\mathbb{P}({\rm SNR_{P}}\geq\beta_t)=P_{\rm active}\mathbb{P}\left(\frac{P_t\|d\|^{-\alpha}h}{\sigma_P^2}\geq\beta_t\right)\nonumber\\
&\stackrel{(a)}{=}P_{\rm active}\exp\left(-\frac{\beta_t\sigma^2_P\|d\|^{\alpha}}{P_t}\right)\nonumber\\
&=\exp\left(-\lambda_e\pi r_g^2-\frac{\beta_t\sigma^2_P\|d\|^{\alpha}}{P_t}\right),
\end{align}
where $\sigma_P^2$ is the noise power at the PT and step (a) follows from $h\sim \exp(1)$. Now we derive secure communication probability for this technique for which we focus on the SNR achieved at the strongest eavesdropper located at $y^{*}$ (as defined in Eq.~\ref{1}), which can be defined as ${\rm SNR_S}=\frac{P_tg_{y^{*}}\|y^{*}\|^{-\alpha}}{\sigma_S^2}$. The secure communication probability is given in the next Lemma.
\begin{lemma}[Secure communication probability]\label{lem:1}
The secure communication probability for the guard zone technique is
\begin{align}
\label{4}
P_{\rm sec}^{GZ}&=\mathbb{P}\left(\frac{P_tg_{y^{*}}\|y^{*}\|^{-\alpha}}{\sigma_S^2}\leq\beta_e\Big|\mathcal{N}(\mathcal{B}(o,r_g))=0\right)\nonumber\\
&=\exp\left(-\frac{2\pi\lambda_e}{\alpha}\left(\frac{P_{t}}{\sigma_S^2\beta_e}\right)^{\frac{2}{\alpha}}\Gamma\left(\frac{2}{\alpha},\frac{r_g^{\alpha}\beta_e\sigma_S^2}{P_{t}}\right)\right),
\end{align}
where $\Gamma(a,b)$ is the upper incomplete gamma function.
\end{lemma}
\begin{IEEEproof}
See Appendix~\ref{app:2}.
\end{IEEEproof}
As evident from Eq.~\ref{3}, $P_{\rm cov}^{GZ}$ is a decreasing function of $r_g$. On the other hand, as noted from Eq.~\ref{4}, the value of $P_{\rm sec}^{GZ}$ is an increasing function of $r_g$. Hence, the optimum value $r_g^{*}$ is the minimum guard zone radius that ensures $P_{\rm sec}^{GZ}\geq\epsilon$. The value of $r_g^{*}$ is derived next.
\begin{lemma}[Optimal guard zone radius]\label{lem:2}
The value of $r_g^{*}$ that maximizes $P_{\rm cov}^{GZ}$ while satisfying the condition of $P_{\rm sec}^{GZ}\geq\epsilon$ is is the one that satisfies the following equation:
\begin{align}
\label{8}
\Gamma\left(\frac{2}{\alpha},\frac{(r_g^*)^{\alpha}\beta_e\sigma_S^2}{P_t}\right)=\min\left\{\frac{\alpha\log\left(\frac{1}{\epsilon}\right)}{2\pi\lambda_{e}\left(\frac{P_{t}}{\sigma_S^2\beta_e}\right)^{\frac{2}{\alpha}}},\Gamma\left(\frac{2}{\alpha}\right)\right\}
\end{align}
\end{lemma}
\begin{IEEEproof}
Substituting the expression of $P_{\rm sec}^{GZ}$ from Eq.~\ref{4} in $P_{\rm sec}^{GZ}\geq\epsilon$, we get $\Gamma\left(\frac{2}{\alpha},\frac{r_g^{\alpha}\beta_e\sigma_S^2}{P_{t}}\right)\leq\frac{\alpha\log\left(\frac{1}{\epsilon}\right)}{2\pi\lambda_{e}\left(\frac{P_{t}}{\sigma_S^2\beta_e}\right)^{\frac{2}{\alpha}}}$. Now if $r_g=0$ satisfies this inequality, then $r_g^*=0$ and $\Gamma\left(\frac{2}{\alpha},\frac{(r_g^*)^{\alpha}\beta_e\sigma_S^2}{P_{t}}\right)=\Gamma\left(\frac{2}{\alpha}\right)$.
Otherwise, the minimum value for $r_g^*$ that satisfies this inequality follows from $\Gamma\left(\frac{2}{\alpha},\frac{(r_g^*)^{\alpha}\beta_e\sigma_S^2}{P_{t}}\right)=\frac{\alpha\log\left(\frac{1}{\epsilon}\right)}{2\pi\lambda_{e}\left(\frac{P_{t}}{\sigma_S^2\beta_e}\right)^{\frac{2}{\alpha}}}$. Combining the results for these two cases leads to the final result in Eq.~\ref{8}.
\end{IEEEproof} 
\vspace{-6mm}
\subsection{Artificial Noise Technique}
\vspace{-2mm}
In this secrecy enhancing technique, the transmission power $P_t$ is split into two parts: (i) $\gamma P_t$, which is used for the transmission of confidential information, and (ii) $(1-\gamma)P_t$, which is used to transmit artificial noise (AN). The AN is generated by using random sequences, which can only be decoded using the keys available at the PRs. Since eavesdroppers do not have access to these keys, they cannot decode AN. Hence, the SNR achieved at the PR is ${\rm SNR_P}=\frac{\gamma P_th\|d\|^{-\alpha}}{\sigma_P^2}$. Please note that unlike the guard zone technique, the PT in this technique is always active, i.e., we have $\delta_a=1$. Hence, the probability of SNR coverage at the PR can be derived as follows:
\begin{align}
\label{5}
P_{\rm cov}^{AN}&=\mathbb{P}({\rm SNR_P}\geq\beta_t)\stackrel{(b)}{=}\exp\left(-\frac{\beta_t\sigma^2_P\|d\|^{\alpha}}{\gamma P_t}\right),
\end{align}
where (b) follows from $h\sim \exp(1)$. On the other hand, we assume that the eavesdropper is unable to decode the AN, which implies that the the SNR at the eavesdropper located at $y^{*}$ is ${\rm SNR_S}=\frac{\gamma P_tg_{y^{*}}\|y^{*}\|^{-\alpha}}{(1-\gamma)P_tg_{y^{*}}\|y^{*}\|^{-\alpha}+\sigma_S^2}$. Hence, the secure communication probability can be derived as follows:
\begin{align}
\label{6}
P_{\rm sec}^{AN}&=\mathbb{P}\left(\frac{\gamma P_tg_{y^{*}}\|y^{*}\|^{-\alpha}}{(1-\gamma)P_tg_{y^{*}}\|y^{*}\|^{-\alpha}+\sigma_S^2}\leq\beta_e\right)\\
&\stackrel{(c)}{=}\mathbb{P}\left(\frac{ {\left(\gamma-(1-\gamma)\beta_e\right)}P_tg_{y^{*}}\|y^{*}\|^{-\alpha}}{\sigma_S^2}\leq{\beta_e}\right)\nonumber,
\end{align}
where step (c) results from simple manipulations of the inequality. This implies that $P_{\rm sec}^{AN}=1$ as long as $\gamma\leq\frac{\beta_e}{1+\beta_e}$. When $\gamma > \frac{\beta_e}{1+\beta_e}$, we can derive a closed-form expression for $P_{\rm sec}^{AN}$ by replacing $\beta_e$ with $\frac{\beta_e}{\gamma-(1-\gamma)\beta_e}$ and $r_g=0$ in Eq.~\ref{4}. This provides the following closed-form expression for $P_{\rm sec}^{AN}$:
\begin{align}
\label{7}
P_{\rm sec}^{AN}=\exp\left(-\frac{2\pi\lambda_e}{\alpha}\left(\frac{P_{t}\left({\gamma-(1-\gamma)\beta_e}\right)}{\sigma_S^2\beta_e}\right)^{\frac{2}{\alpha}}\Gamma\left(\frac{2}{\alpha}\right)\right).
\end{align}

Since the power used for information transmission is directly proportional to $\gamma$, $P_{\rm cov}^{AN}$ is an increasing function of $\gamma$, which is evident from Eq.~\ref{5}. On the other hand, since the power used for transmitting AN is directly proportional to $1-\gamma$, $P_{\rm sec}^{AN}$ decreases with increase in $\gamma$, which is evident from Eq.~\ref{7}. Therefore, the optimum value of $\gamma^{*}$ is the maximum value of $\gamma$ that ensures $P_{\rm sec}^{AN}\geq\epsilon$. This optimal $\gamma^{*}$ is derived in the following Lemma. 
\begin{lemma}[Optimal power split]\label{lem:3}
The value of $\gamma^{*}$ that maximizes $P_{\rm cov}^{AN}$ while satisfying the condition $P_{\rm sec}^{AN}\geq\epsilon$ is 
\begin{align}
\label{9}
\gamma^{*}=\min\left\{1,\frac{\beta_e}{1+\beta_e}\left(1+\frac{\sigma_S^2}{P_t}\left(\frac{\alpha \log\left(\frac{1}{\epsilon}\right)}{2\pi\lambda_e\Gamma(\frac{2}{\alpha})}\right)^{\frac{\alpha}{2}}\right)\right\}.
\end{align}
\end{lemma}
\begin{IEEEproof}
The result follows by substituting Eq.~\ref{7} in the inequality $P_{\rm sec}^{AN}\geq\epsilon$, and following similar approach as in the proof of Lemma~\ref{lem:2}.
\end{IEEEproof}
\vspace{-4mm} 
\section{Performance Comparison}
\subsection{Useful Threshold on the Density of Eavesdroppers}
In this subsection, we first aim to find the threshold on $\lambda_e$ below which the secrecy enhancing techniques are not required ($r_g^{*}=0$ and $\gamma^{*}=1$). Note that when $r_g^{*}=0$ and $\gamma^{*}=1$ the performance is the same for both techniques: $P_{\rm cov}^{GZ}=P_{\rm cov}^{AN}$ and $P_{\rm sec}^{GZ}=P_{\rm sec}^{AN}$. For the guard zone technique, we can derive this threshold by solving the following inequality:
\begin{align}
\label{10}
\Gamma\left(\frac{2}{\alpha}\right)\leq\frac{\alpha\log\left(\frac{1}{\epsilon}\right)}{2\pi\lambda_{e}\left(\frac{P_{t}}{\sigma_S^2\beta_e}\right)^{\frac{2}{\alpha}}},
\end{align}
where this inequality ensures that the result of Eq.~\ref{8} is $r_g^{*}=0$. Solving this inequality, we deduce that $r_g^{*}=0$ as long as
\begin{align}
\label{11}
\lambda_e\leq \frac{\alpha}{2\pi \Gamma\left(\frac{2}{\alpha}\right)} \log\left(\frac{1}{\epsilon}\right)
\left(\frac{P_{t}}{\sigma_S^2\beta_e}\right)^{-\frac{2}{\alpha}}.
\end{align} 
Similarly, for the artificial noise technique, we can derive this threshold  on $\lambda_e$ by solving the following inequality:
\begin{align}
\label{12}
\frac{\beta_e}{1+\beta_e}\left(1+\frac{\sigma_S^2}{P_t}\left(\frac{\alpha \log\left(\frac{1}{\epsilon}\right)}{2\pi\lambda_e\Gamma(\frac{2}{\alpha})}\right)^{\frac{\alpha}{2}}\right) \geq 1,
\end{align}
where the above inequality ensures that the result of Eq.~\ref{9} is $\gamma^{*}=1$. Solving the above inequality, we infer that artificial noise addition is not required as long as
\begin{align}
\label{13}
\lambda_e\leq \frac{\alpha}{2\pi \Gamma\left(\frac{2}{\alpha}\right)} \log\left(\frac{1}{\epsilon}\right)
\left(\frac{P_{t}}{\sigma_S^2\beta_e}\right)^{-\frac{2}{\alpha}}.
\end{align}
As can be expected intuitively, the threshold on $\lambda_e$ derived in Eq.~\ref{11} and Eq.~\ref{13} for the two techniques is the same. We denote this threshold by $\lambda_e^{*}$.  
As long as $\lambda_e < \lambda_e^{*}$, the secure communication probability is guaranteed to be above $\epsilon$. 
\vspace{-5mm}	
\subsection{Comparison of Secrecy Enhancement Techniques}
In this subsection, we focus on $\lambda_e\geq\lambda_e^{*}$ for which secrecy enhancement techniques are required to ensure desired secrecy performance level. In particular, we will characterize regimes in which a given technique outperforms the other. Since both techniques select their parameters ($r_g^{*}$ or $\gamma^{*}$) in order to ensure that $P_{\rm sec}\geq\epsilon$, optimal parameter choices will naturally satisfy the desired secrecy conditions. As a result, we focus our comparison on the other system performance metric: $P_{\rm cov}^{GZ}$ and $P_{\rm cov}^{AN}$. Hence, for $\lambda_e\geq\lambda_e^{*}$, a given secrecy technique is said to perform better than the other if it provides higher coverage probability at the PR while ensuring $P_{\rm sec}\geq\epsilon$. 
In the following theorem, we characterize the regimes in which a given secrecy enhancing technique outperforms the other. 
\vspace{-2mm}
\begin{theorem}[Secrecy enhancement technique selection] \label{thm:1}
Defining the functions $\mathcal{F}$, $\mathcal{H}$, and $\mathcal{G}$ as
\begin{align}
\label{14}
\mathcal{F}&=\frac{\alpha\log\left(\frac{1}{\epsilon}\right)}{2\pi\lambda_e\left(\frac{P_t}{\sigma_S^2\beta_e}\right)^{\frac{2}{\alpha}}}-\Gamma\left(\frac{2}{\alpha},\mathcal{H}\right)\\
\mathcal{H}&=\frac{\beta_e\sigma_S^2}{P_t}\left[\frac{\beta_t d^{\alpha}\sigma_S^2}{P_t\lambda_e\pi}\left(\frac{1}{\mathcal{G}}-1\right)\right]^{\frac{\alpha}{2}}\\
\mathcal{G}&=\frac{\beta_e}{1+\beta_e}\left(1+\frac{\sigma_S^2}{P_t}\left(\frac{\alpha \log\left(\frac{1}{\epsilon}\right)}{2\pi\lambda_e\Gamma(\frac{2}{\alpha})}\right)^{\frac{\alpha}{2}}\right),
\end{align}
the guard zone technique is a better choice as long as $\mathcal{F}> 0$, while artificial noise technique is a better choice when $\mathcal{F}\leq 0$.
\end{theorem}
\begin{IEEEproof}
See Appendix~\ref{app:1}.
\end{IEEEproof}
\begin{remark} \label{rem:d}
Observing the dependence of $\mathcal{F}$ on D2D link distance $d$ in Eq.~\ref{14}, we note that the value of $d$ plays an important role in determining which technique performs better. Since $\mathcal{F}$ is an increasing function of $d$, it is easy to conclude that for a given set of system parameters, the artificial noise technique provides better performance at lower values of $d$, while the guard zone technique starts performing better when $d$ exceeds a specific threshold. These comments will be verified next in the numerical results section.
\end{remark}
\vspace{-5mm}
\subsection{Numerical Results} 
\vspace{-1mm}
\begin{figure}
\centering
\includegraphics[width=0.47\columnwidth]{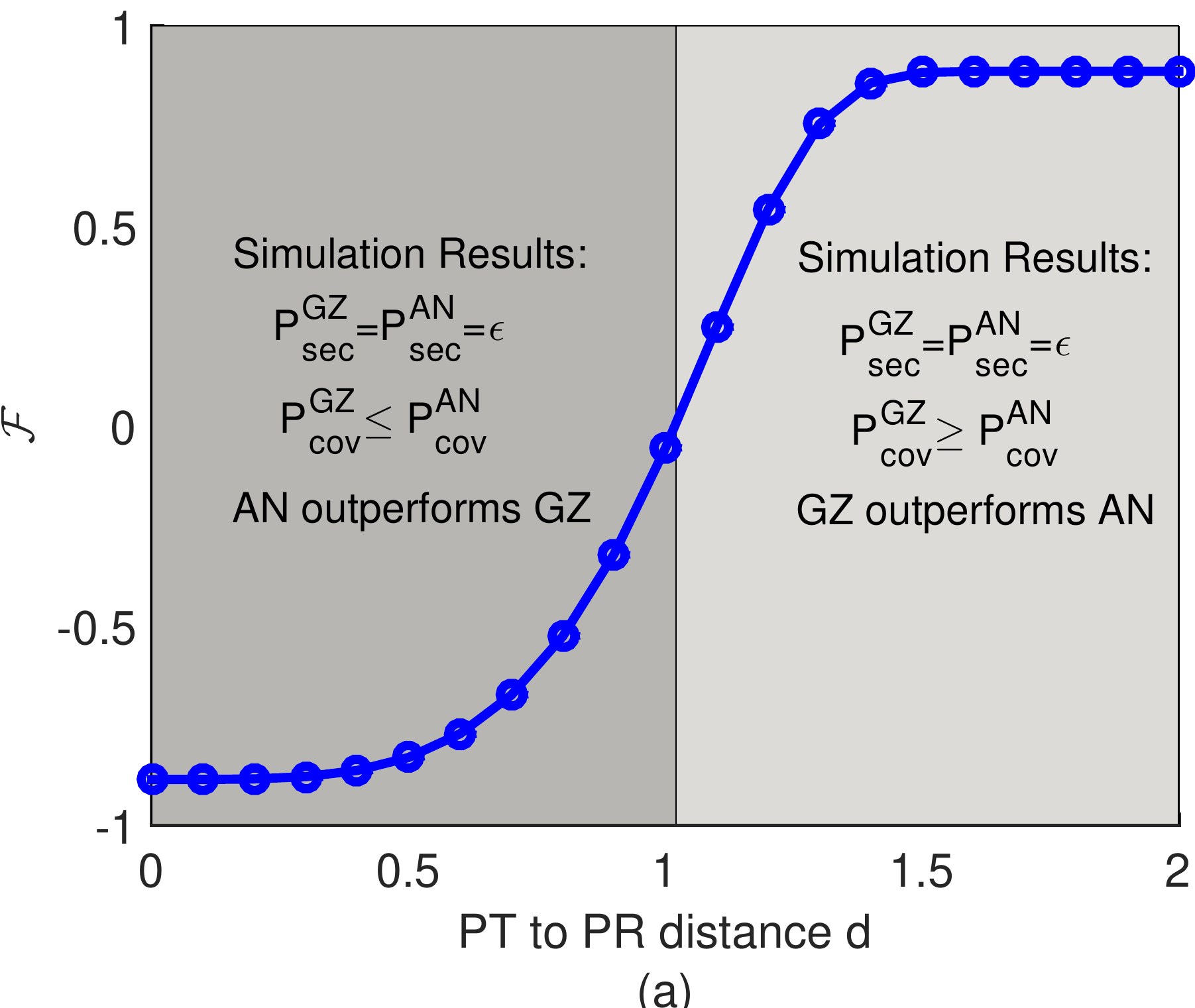}
\hfill
\includegraphics[width=0.47\columnwidth]{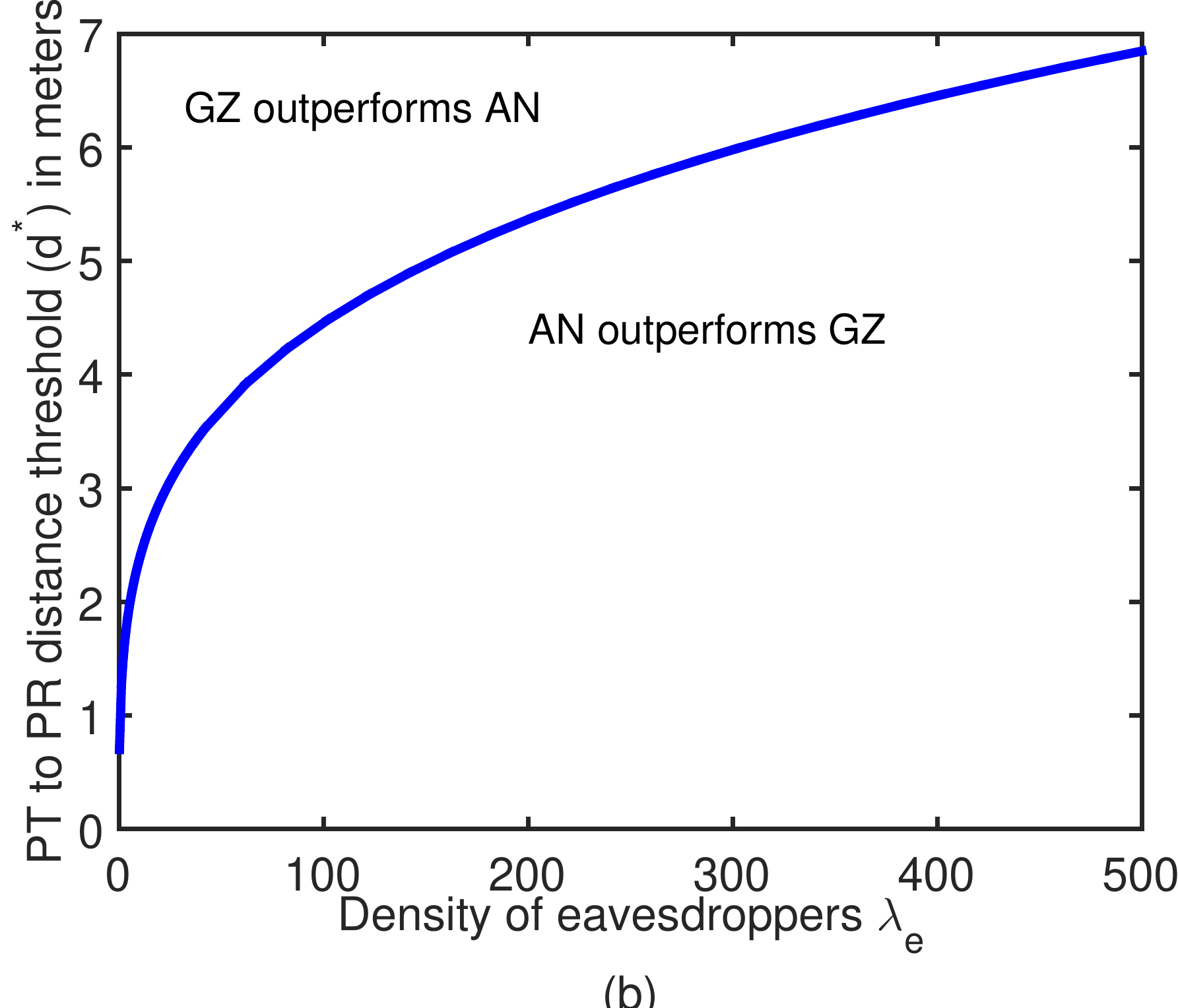}
\caption{(a) Technique selection function $\mathcal{F}$ as a function of $d$, and (b) PT to PR distance threshold $d^*$ for different values of $\lambda_e$.}
\label{fig:1}
\vspace{-5mm}
\end{figure}
For numerical comparisons, we consider the following system parameters: $\alpha=4$, $P_t=1$, $\beta_t=2$, $\beta_e=1$, $\epsilon=0.9$, $\sigma_P^2=1$, and $\sigma_S^2=1$. For this setup, $\lambda_e^* = 0.0378$ because of which we choose $\lambda_e = 0.1 > \lambda_e^*$. In Fig.~\ref{fig:1}.a, we use Monte-Carlo simulations to evaluate the values of $P_{\rm sec}^{GZ}$, $P_{\rm sec}^{AN}$, $P_{\rm cov}^{GZ}$, and $P_{\rm cov}^{AN}$ to determine which technique is better at different values of $d$. On the same figure, we plot the function $\mathcal{F}$ derived in Theorem~\ref{thm:1}. The comparison of the simulation and analytical results supports the main consequence of our analysis that the guard zone technique is a better choice when $\mathcal{F}> 0$ while the artificial noise technique is a better choice when $\mathcal{F}\leq 0$. In addition, our comment in Remark~\ref{rem:d} that artificial noise technique is better for lower values of $d$ is verified. It is clear from this comparison that the value of $d$ at which $\mathcal{F}$ switches from being negative to positive is critical to the choice of the secrecy enhancement technique. We refer to this threshold value of $d$ as $d^*$. In Fig.~\ref{fig:1}.b, we study the effect of $\lambda_e$ on $d^*$. The resulting curve partitions the $(d^*, \lambda_e)$ plane into two parts: lower part in which AN outperforms GZ and the the upper part in which GZ outperforms AN. We notice that with increasing $\lambda_e$, $d^*$ increases, which means AN starts becoming optimal choice for a larger range of values for $d$. 
\vspace{-4mm}
\section{Conclusion}
\vspace{-2mm}
In this letter, we provided a concrete performance comparison of two popular secrecy enhancement techniques: (i) creating guard zones around legitimate transmitters, and (ii) adding artificial noise to the confidential messages. Using tools from stochastic geometry, we first derived a closed-form expression for the threshold on the density of eavesdroppers below which no secrecy enhancement techniques are required. For densities greater than this threshold, we characterized regimes in which a given secrecy enhancement technique outperforms the other. Our results demonstrate that guard zone technique is a better choice when the distances between the D2D pairs are higher than a specific threshold. 

A key technical extension for this line of work is the inclusion of interference in the analysis. This requires a significantly more complicated analysis due to spatial correlation between interference levels at the PR and the eavesdropper which requires joint analysis of coverage probability at the PR and secure communication probability at the eavesdropper.
\vspace{-2mm}
\appendix
\vspace{-3mm}
\subsection{Proof of Lemma~\ref{lem:1}}\label{app:2}
By definition, the secure communication probability is
\vspace{-3mm}
\begin{align}
P_{\rm sec}^{GZ}&=\mathbb{P}\left(\frac{P_{t}}{\sigma_S^2}\underset{y\in{\Phi}_{e}\cap\mathcal{B}(0,r_g)^c}{\max}{\{g_y\|y\|^{-\alpha}\}}\leq\beta_e\right)\nonumber\\
&\stackrel{(a)}{=}\mathbb{E}_{{\Phi}_{e}}\left[\prod_{y\in{\Phi}_{e}\cap\mathcal{B}(0,r_g)^c}\mathbb{P}\left(g_y\leq\|y\|^{\alpha}\beta_e\frac{\sigma_S^2}{P_{t}}\Big|\Phi_e\right)\right]\nonumber\\
&\stackrel{(b)}{=}\mathbb{E}_{{\Phi}_{e}}\left[\prod_{y\in{\Phi}_{e}\cap\mathcal{B}(0,r_g)^c}\left(1-e^{-\|y\|^{\alpha}\beta_e\frac{\sigma_S^2}{P_{t}}}\right)\right]\nonumber\\
&\stackrel{(c)}{=}\exp\left(-2\pi{\lambda}_{e}\int_{r_g}^\infty e^{-r_y^{\alpha}\beta_e\frac{\sigma_S^2}{P_{t}}}r_y{\rm d}r_y\right),
\end{align}
where $\mathcal{B}(0,r_g)^c$ is the compliment of the area covered by the ball centered at the origin with radius $r_g$. Step (a) follows from the independence of $g_y$ across all wireless links, (b) is due to $g_y\sim \exp(1)$, (c) follows from applying PGFL of PPP and converting to polar coordinates. With simple algebraic manipulations, the final result presented in Lemma~\ref{lem:1} follows.%
\vspace{-4mm}
\subsection{Proof of Theorem~\ref{thm:1}}\label{app:1}
Technique selection is done using the following inequality:
\begin{align}
\label{17}
&P_{\rm cov}^{GZ}\underset{AN}{\overset{GZ}{\gtrless}}P_{\rm cov}^{AN}\nonumber\\
\Rightarrow&\ \exp\left(-\lambda_e\pi {r_g^{*}}^2-\frac{\beta_t\sigma^2_P\|d\|^{\alpha}}{P_t}\right)\underset{AN}{\overset{GZ}{\gtrless}}\exp\left(-\frac{\beta_t\sigma^2_P\|d\|^{\alpha}}{\gamma^{*} P_t}\right)\nonumber\\
\Rightarrow&\ \frac{\beta_t\sigma^2_P\|d\|^{\alpha}}{\gamma^{*} P_t}\underset{AN}{\overset{GZ}{\gtrless}} \lambda_e\pi {r_g^{*}}^2+\frac{\beta_t\sigma^2_P\|d\|^{\alpha}}{P_t}\nonumber\\ 
\Rightarrow&\ \frac{\beta_t\sigma^2_P\|d\|^{\alpha}}{P_t}\left(\frac{1}{\gamma^{*}}-1\right)\underset{AN}{\overset{GZ}{\gtrless}} \lambda_e\pi {r_g^{*}}^2 
\end{align}
Since $\lambda_e\geq\lambda_e^{*}$, then $\Gamma\left(\frac{2}{\alpha},\frac{(r_g^*)^{\alpha}\beta_e\sigma_S^2}{P_t}\right)=\frac{\alpha\log\left(\frac{1}{\epsilon}\right)}{2\pi\lambda_{e}\left(\frac{P_{t}}{\sigma_S^2\beta_e}\right)^{\frac{2}{\alpha}}}$ and $\gamma^{*}=\mathcal{G}=\frac{\beta_e}{1+\beta_e}\left(1+\frac{\sigma_S^2}{P_t}\left(\frac{\alpha \log\left(\frac{1}{\epsilon}\right)}{2\pi\lambda_e\Gamma(\frac{2}{\alpha})}\right)^{\frac{\alpha}{2}}\right)$. Substituting this in Eq.~\ref{17}, we get: 
\begin{align}
\label{18}
&\frac{\beta_e\sigma_S^2}{P_t}\left(\frac{\beta_t\sigma^2_P\|d\|^{\alpha}}{\lambda_e\pi P_t}\left(\frac{1}{\gamma^{*}}-1\right)\right)^{\frac{\alpha}{2}}\underset{AN}{\overset{GZ}{\gtrless}} \frac{(r_g^*)^{\alpha}\beta_e\sigma_S^2}{P_t}\nonumber\\
\Rightarrow&\ \Gamma\left(\frac{2}{\alpha},\frac{(r_g^*)^{\alpha}\beta_e\sigma_S^2}{P_t}\right) \underset{AN}{\overset{GZ}{\gtrless}}\Gamma\Bigg(\frac{2}{\alpha},\mathcal{H}\Bigg)\\
\Rightarrow&\ \frac{\alpha\log\left(\frac{1}{\epsilon}\right)}{2\pi\lambda_{e}\left(\frac{P_{t}}{\sigma_S^2\beta_e}\right)^{\frac{2}{\alpha}}} \underset{AN}{\overset{GZ}{\gtrless}}\Gamma\Bigg(\frac{2}{\alpha},\mathcal{H}\Bigg)\Rightarrow \mathcal{F} \underset{AN}{\overset{GZ}{\gtrless}} 0,
\end{align}
where (21) follows by substituting $\gamma^{*}=\mathcal{G}$ and substituting for $\mathcal{H}$ as defined in Theorem~\ref{thm:1} and taking $\Gamma$ on both sides, while (22) follows by substituting for $\mathcal{F}$ as defined in Theorem~\ref{thm:1}. This concludes the proof.
\vspace{-3mm}
\bibliographystyle{IEEEtran}
\bibliography{Dhillon_WCL2017-0001}

\begin{thebibliography}{10}
\providecommand{\url}[1]{#1}
\csname url@samestyle\endcsname
\providecommand{\newblock}{\relax}
\providecommand{\bibinfo}[2]{#2}
\providecommand{\BIBentrySTDinterwordspacing}{\spaceskip=0pt\relax}
\providecommand{\BIBentryALTinterwordstretchfactor}{4}
\providecommand{\BIBentryALTinterwordspacing}{\spaceskip=\fontdimen2\font plus
\BIBentryALTinterwordstretchfactor\fontdimen3\font minus
  \fontdimen4\font\relax}
\providecommand{\BIBforeignlanguage}[2]{{%
\expandafter\ifx\csname l@#1\endcsname\relax
\typeout{** WARNING: IEEEtran.bst: No hyphenation pattern has been}%
\typeout{** loaded for the language `#1'. Using the pattern for}%
\typeout{** the default language instead.}%
\else
\language=\csname l@#1\endcsname
\fi
#2}}
\providecommand{\BIBdecl}{\relax}
\BIBdecl

\bibitem{5934342}
X.~Zhou, R.~K. Ganti, J.~G. Andrews, and A.~Hjorungnes, ``On the throughput
  cost of physical layer security in decentralized wireless networks,''
  \emph{IEEE Trans. on Wireless Commun.}, vol.~10, no.~8, pp. 2764--2775, Aug.
  2011.

\bibitem{zhou2010secure}
X.~Zhou and M.~R. McKay, ``Secure transmission with artificial noise over
  fading channels: Achievable rate and optimal power allocation,'' \emph{IEEE
  Trans. on Veh. Technology}, vol.~59, no.~8, pp. 3831--3842, Oct. 2010.

\bibitem{7247765}
W.~Liu, Z.~Ding, T.~Ratnarajah, and J.~Xue, ``On ergodic secrecy capacity of
  random wireless networks with protected zones,'' \emph{IEEE Trans. on Veh.
  Technology}, vol.~65, no.~8, pp. 6146--6158, Aug. 2016.

\bibitem{6259796}
J.~Xiong, K.~K. Wong, D.~Ma, and J.~Wei, ``A closed-form power allocation for
  minimizing secrecy outage probability for {MISO} wiretap channels via masked
  beamforming,'' \emph{IEEE Commun. Letters}, vol.~16, no.~9, pp. 1496--1499,
  Sept. 2012.

\bibitem{AndGupJ2016}
J.~G. Andrews, A.~K. Gupta, and H.~S. Dhillon, ``A primer on cellular network
  analysis using stochastic geometry,'' 2016, available online:
  arxiv.org/abs/1604.03183.

\bibitem{7073589}
H.~Sun, M.~Wildemeersch, M.~Sheng, and T.~Q.~S. Quek, ``{D2D} enhanced
  heterogeneous cellular networks with dynamic {TDD},'' \emph{IEEE Trans. on
  Wireless Commun.}, vol.~14, no.~8, pp. 4204--4218, Aug. 2015.

\bibitem{7056528}
A.~H. Sakr and E.~Hossain, ``Cognitive and energy harvesting-based {D2D}
  communication in cellular networks: Stochastic geometry modeling and
  analysis,'' \emph{IEEE Trans. on Commun.}, vol.~63, no.~5, pp. 1867--1880,
  May 2015.

\bibitem{7446343}
M.~Afshang, H.~S. Dhillon, and P.~H.~J. Chong, ``Modeling and performance
  analysis of clustered device-to-device networks,'' \emph{IEEE Trans. on
  Wireless Commun.}, vol.~15, no.~7, pp. 4957--4972, July 2016.

\bibitem{geraci2014physical}
G.~Geraci, H.~S. Dhillon, J.~G. Andrews, J.~Yuan, and I.~B. Collings,
  ``Physical layer security in downlink multi-antenna cellular networks,''
  \emph{IEEE Trans. on Commun.}, vol.~62, no.~6, pp. 2006--2021, June 2014.

\bibitem{6772207}
A.~D. Wyner, ``The wire-tap channel,'' \emph{The Bell System Technical
  Journal}, vol.~54, no.~8, pp. 1355--1387, Oct. 1975.

\bibitem{4802331}
X.~Tang, R.~Liu, P.~Spasojevic, and H.~V. Poor, ``On the throughput of secure
  hybrid-{ARQ} protocols for gaussian block-fading channels,'' \emph{IEEE
  Trans. on Info. Theory}, vol.~55, no.~4, pp. 1575--1591, Apr. 2009.

\bibitem{5701754}
X.~Zhou, M.~R. McKay, B.~Maham, and A.~Hjorungnes, ``Rethinking the secrecy
  outage formulation: A secure transmission design perspective,'' \emph{IEEE
  Commun. Letters}, vol.~15, no.~3, pp. 302--304, Mar. 2011.

\end{thebibliography}

\end{document}